\documentclass[12pt]{JHEP3}
\usepackage{amsmath}
\usepackage{epsf}
\usepackage{epsfig}
\usepackage{here}
\usepackage{amssymb}
\usepackage{citesort}
\usepackage{graphicx}
\usepackage{latexsym}
\input axodraw.sty
\input epsf.sty

\newcommand{\bce}{\begin{center}}
\newcommand{\ece}{\end{center}}
\newcommand{\bi}{\begin{itemize}}
\newcommand{\ei}{\end{itemize}}
\newcommand{\bq}{\begin{equation}}
\newcommand{\eq}{\end{equation}}
\newcommand{\bqa}{\begin{eqnarray}}
\newcommand{\eqa}{\end{eqnarray}}

\newcommand{\bt}{\begin{tabbing}}
\newcommand{\et}{\end{tabbing}}
\newcommand{\nn}{\nonumber}
\newcommand{\be}{\begin{equation}}
\newcommand{\ee}{\end{equation}}
\newcommand{\bear}{\begin{eqnarray}}
\newcommand{\ear}{\end{eqnarray}}

\newsavebox{\LSIM}
\sbox{\LSIM}{\raisebox{-1ex}{$\ \stackrel{\textstyle<}{\sim}\ $}}
\newcommand{\lsim}{\usebox{\LSIM}}
\newsavebox{\GSIM}
\sbox{\GSIM}{\raisebox{-1ex}{$\ \stackrel{\textstyle>}{\sim}\ $}}
\newcommand{\gsim}{\usebox{\GSIM}}
\renewcommand{\vec}[1]{{\bf #1}}

\title{The Baryon asymmetry in the Standard Model with a low cut-off}

\author{
Dietrich B\"odeker${}^1$, Lars Fromme${}^1$, Stephan J.~Huber${}^2$ 
and Michael Seniuch${}^1$\\
${}^1$Fakult\"at f\"ur Physik, Universit\"at Bielefeld, D-33615 Bielefeld, Germany \\
${}^2$Theory Division, Physics Department, CERN, CH-1211 Geneva 23, Switzerland
\\
Emails:\\ \email{bodeker@physik.uni-bielefeld.de},
       \email{fromme@physik.uni-bielefeld.de},
       \email{stephan.huber@cern.ch}, \email{seniuch@physik.uni-bielefeld.de}}

\abstract{We study the generation of the baryon asymmetry in a variant of the
standard model, where the Higgs field is stabilized by a dimension-six 
interaction. Analyzing the one-loop potential, we find a strong 
first order electroweak phase transition for Higgs masses up to at least 
170 GeV. Dimension-six operators induce also new sources of CP violation.
We compute the baryon asymmetry in the WKB approximation.
Novel source terms in the transport equations enhance the generated
baryon asymmetry. For a wide range of parameters the model predicts 
a baryon asymmetry close to the observed value.
}

\keywords{baryogenesis, electroweak phase transition, higher dimensional operators}

\preprint{BI-TP 2004/41, CERN-PH-TH/2004/258}

\begin{document}
 

%
%
%
%
\section{Introduction}
The baryon asymmetry of the universe has recently been determined to
an unprecedented accuracy, 
\begin{equation} 
\label{eta}
\eta_B\equiv \frac{n_B}{s}=(8.9\pm0.4)\times10^{-11},
\end{equation} 
by combining measurements of the cosmic microwave background \cite{CMB}
and large scale structures \cite{LSS}. Explaining its origin is one of the great 
challenges of modern particle physics and cosmology. For baryogenesis Sakharov's
three conditions, B violation, CP violation and deviation from 
thermal equilibrium have
to be satisfied. In principle these conditions could be met within the standard 
model (SM) 
at the electroweak phase transition (EWPT) \cite{Kuzmin}. 
A more quantitative analysis shows 
however that the baryon asymmetry cannot be explained within the SM because 
there is not enough CP violation \cite{FS93} and the phase transition turns into a 
smooth crossover for Higgs masses $m_H\gsim$80 GeV \cite{KLRS96}. In fact,
electroweak baryogenesis requires an even stronger criterion to be satisfied: 
The Higgs vacuum expectation value (vev) at the critical temperature, 
$v_c\equiv\langle\phi(T_c)\rangle$,
must be larger than about $T_c$ in order to avoid baryon number washout
after the phase transition.  

Motivated by the possibility that electroweak baryogenesis can be tested at
future colliders,  there were many proposals to realize this mechanism
in extended settings \cite{review}. Some recent attempts can be found
in ref.~\cite{new}. In supersymmetric
models a strong EWPT can be induced by a light top squark \cite{MSSM}. Supersymmetry
breaking also provides new sources of CP violation. However, by now this scenario is 
quite constrained due to the negative Higgs searches. In the SM 
a lower bound of $m_H>$114 GeV was established \cite{m_H}.  
A strong first order phase transition could also be driven by cubic interactions 
of a singlet Higgs field \cite{NMSSM}.

Recently, an alternative idea caught attention: non-renormalizable operators could
have an impact on the EWPT. These operators parametrize the effects of new 
physics beyond some cut-off scale $M$. In order to be relevant at weak 
scale temperatures we have to assume that $M\lsim1$ TeV.
This new dynamics could be an ordinary quantum field theory, 
e.g.~an extended Higgs sector. It might as well be something more fundamental,
like strongly coupled gravity if the hierarchy problem
is solved by the presence of extra dimensions. 

If the Higgs potential is stabilized by a $\phi^6$ interaction, a strong first 
order phase transition can occur for Higgs masses well above 100 GeV 
\cite{Z93,GSW04,HO04}. A first order transition is triggered by a barrier in
the Higgs potential. It can be provided by the one-loop thermal corrections 
of the weak gauge bosons. In the model under consideration, a 
barrier can also be generated from a negative
$\phi^4$ term, which no longer destabilizes the Higgs potential.
The latter possibility turns out to be dominant in a large part of the
parameter space. 
Non-renormalizable interactions also allow for new sources of the 
CP violation to fuel baryogenesis \cite{DHSS91,ZLWY94}. 

In the following we will investigate the strengh of the EWPT in the SM with
low cut-off, taking into account the one-loop corrections to the potential.
At one-loop the phase transition is somewhat weaker than found in the
analysis of ref.~\cite{GSW04}, where only the thermal mass part of the
one-loop correction was taken into account.
Still we find a strong 
first order EWPT for Higgs masses up to 
at least 170 GeV, if we require $M>500$ GeV.  We will study the
properties of the bubble profile, finding in particular that the
wall thickness varies in a wide range from about 2 to 16 times $1/T_c$.
We will discuss dimension-6 interactions between the Higgs field 
and the top quark which provide the necessary CP violation to fuel
baryogenesis. In the WKB approximation these operators induce
CP violating terms in the top quark dispersion relation which vary
along the bubble wall and enter the transport equations as source terms. 
We will discuss novel source terms in the transport equations which
enhance the generated baryon asymmetry. We find that the model
can explain the observed baryon asymmetry for natural
values of the parameters. 

%
%
\section{The strength of the phase transition}
The dynamics of the EWPT is determined by the effective Higgs potential.
As proposed in ref.~\cite{Z93}, we add a non-renormalizable
$\phi^6$ operator to the SM potential, so that
\be
V(\phi)=-{\mu^2 \over 2}\phi^2+{\lambda \over 4}\phi^4+{1\over
  {8M^2}}\phi^6
  ,
\ee
where $ \phi^2  \equiv 2\Phi^{\dagger}\Phi  $ with the 
 SM Higgs doublet $\Phi$. 

At finite temperature we add a thermal mass term to the potential. Because
of the positive definite $\phi^6$-term, the quartic coupling $\lambda$ is allowed to
take
negative values. In the high temperature
expansion of the one-loop thermal potential we get the thermal Higgs mass term 
\be \label{thermmass}
{1\over 2}\left({1\over 2}\lambda+{3\over 16}g_2^2+{1\over 16}{g_1}^2+{1\over
    4}y _ t ^2\right)T^2 \phi^2,
\ee
where $y _ t $ is the top Yukawa coupling and $g_2$ and $g_1$ are the $SU(2)_L$ and
$U(1)_Y$ gauge couplings. 
We also include the one-loop contributions due to the transverse 
gauge bosons

\be \label{cubic}
-{{g^3_2}\over{16\pi}}T\phi^3
\ee
and the top quark
\be \label{log}
{3\over{64\pi^2}}y _ t ^4 \phi^4 \ln \left({Q^2 \over {c_F T^2}}\right).
\ee 
to the effective potential, where $c_F\approx13.94$ \cite{DJ74}. 
We choose $Q=m_{\rm top}=178$ GeV.
Another choice of $Q$ would only change the value of the self-coupling
$\lambda$. 
Moreover we add the leading one-loop and two-loop corrections due to the
$\phi^6$ interaction
\be \label{1_2loop}
{1\over{8M^2}}(2\phi^4 T^2+ \phi^2 T^4).
\ee
Altogether we end up with the high temperature effective potential
\bqa
V_{\rm eff}(\phi,T)=&&{1\over 2}\left(-\mu^2+\left({1\over 2}\lambda+{3\over
      16}g_2^2+{1\over 16}{g_1}^2+{1\over 4}y _ t ^2\right)T^2\right)
\phi^2 \nn\\
&-&{{g^3_2}\over{16\pi}}T\phi^3 +{\lambda \over
  4}\phi^4+{3\over{64\pi^2}}y _ t ^4 \phi^4 \ln \left({Q^2 \over {c_F
      T^2}}\right)\nn\\
&+&{1\over{8M^2}}(\phi^6+2\phi^4 T^2+ \phi^2 T^4).\label{Veff}
\eqa

With the two conditions 
\be
\left.{\partial V_{\rm eff}(\phi,0)\over \partial \phi}\right|_{\phi=v}=0
\mbox{\hspace{1.0cm}and\hspace{1.0cm}} {\partial^2 V_{\rm eff}(\phi,0)\over \partial
  \phi^2}=m_H^2,
\ee
where 
\be
  V_{\rm eff}(\phi,0)=  V ( \phi  ) 
  -{3\over{64\pi^2}}y _ t ^4 \phi^4 \ln \left({y _ t ^2 \phi^2\over
    2Q^2}\right)
\ee
is the zero-temperature potential including the one-loop correction from 
the top-quark,  we can express the two parameters $\mu$ and $\lambda$ by the
physical quantities $m_H$ and $v=246$ GeV. 
In the following we take $m_H$ and $M$ as the free parameters of
the model. The SM bound on the Higgs mass applies to our model, so we 
require $m_H>114$ GeV. We need $M\lsim 1$ TeV in order for the dimension-six operator
to be of relevance for the phase transition. If $M$ becomes too small, for
a fixed value of $m_H$, the electroweak minimum ceases to be
the global minimum of the zero-temperature potential. As shown
in fig.~\ref{xi-lines}, this happens around $M<500$ GeV, and we
exclude these values from the parameter space.

During a first order phase transition
there exist two energetically degenerate phases separated by an energy barrier
at the critical temperature $T_c$. To obtain $T_c$ and the non-zero value of
the vacuum expectation value $v_c$ the two conditions
\be
\left.{\partial V_{\rm eff}(\phi,T_c)\over \partial \phi}\right|_{\phi=v_c}=0
\mbox{\hspace{1.0cm}and\hspace{1.0cm}} V_{\rm eff}(v_c,T_c)=0
\ee
have to be fulfilled.
\FIGURE[t]{
   \epsfig{file=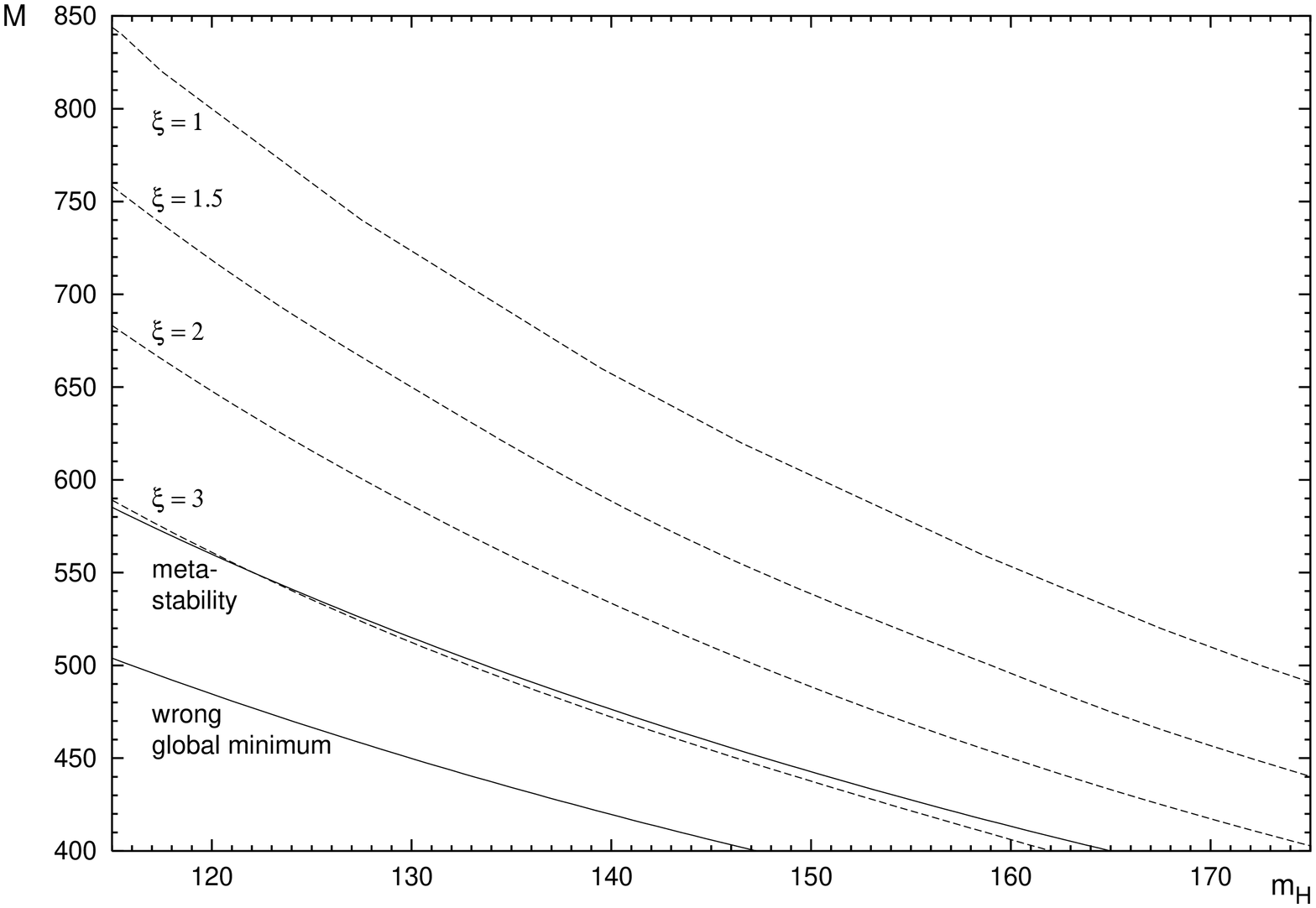,width=80mm,angle=270}
\caption{Contours of constant $\xi=v_c/T_c$ in the $M$-$m_H$-plane. $M$ and $m_H$
         are given in units of GeV.  Below the lowest line the
         zero-temperature minimum at $\phi\ne 0$ is no longer the global
         one. Below the metastability line the probability for thermal tunneling
         gets too small.}
\label{xi-lines}
}
The critical temperature in case of the EWPT is around 100 GeV.
At some particular temperature below $T_c$, say $T_n$ (nucleation temperature),
the broken phase bubbles nucleate, expand and percolate. 
The Higgs field changes
rapidly as the bubble wall passes through space. Baryogenesis has to
take place outside the bubble while within the bubble the sphaleron induced
$(B+L)$-violating reactions must be strongly suppressed. Otherwise the generated
baryon asymmetry would be washed out after the phase transition. The sphaleron
rate is practically switched off if the "washout criterion" \cite{mooreBrokenPhase} 
\be \label{wo}
\xi={v_c\over T_c} \ge 1.1
\ee
is satisfied. This is the condition for a first order transition to be
strong.  As was discussed in ref.~\cite{GSW04}, the sphaleron energy and
therefore eq.~(\ref{wo}) are practically not affected by the presence of
the $\phi^6$ term.

In fig.~\ref{xi-lines} we show the strength of the phase transition as a
function of the model parameters. As expected the EWPT becomes
weaker for increasing Higgs masses. For the smallest allowed Higgs mass
we need $M\lsim$ 825 GeV to satisfy the washout criterion.  In contrast
to the SM we find a strongly first order phase transition, even for
Higgs masses above 150 GeV. A large part of the parameter space meets the
requirements of electroweak baryogenesis.
As $M$ approaches the region of wrong zero-temperature minimum,
the critical temperature becomes smaller and $\xi$ larger.
For $\xi\gsim3$ the high temperature approximation breaks down for the top quark.

What Higgs masses are compatible with the washout criterion depends 
on how small $M$ is allowed to be. There is no particular bound on the
$\phi^6$ operator \cite{BHLMZ03}. However, in fig.~\ref{xi-lines} we take
$M\gsim 400$ GeV
in order to make an expansion in powers of $v/M$ reasonable. In an
effective field theory all operators which are allowed by the symmetries
are expected to be present. In particular, we expect dimension-6 operators
involving gauge fields, such as $(1/M^2)(\Phi^{\dagger}D_{\mu}\Phi)^2$.
These operators have to be suppressed by a higher scale of about 10 TeV in order
to be in agreement with the electroweak precision data \cite{GSW04}. 
Thus a tuning of couplings on the order of (10 TeV/$M)^2$ is required,
and has to be explained by the UV completion of the model.

At the one-loop level the phase transition is somewhat weaker
compared to the analysis of ref.~\cite{GSW04}. There only the
thermal masses (\ref{thermmass}) were included in the computation.
For instance, taking $m_H=115$ GeV, we find $M=825$ GeV to arrive
at a strong enough EWPT, i.e.~$\xi=1.1$. Including only the thermal
mass corrections, one arrives at $\xi=1.43$, and the cut-off scale
can be increased to about 870 GeV until the phase transition
becomes too weak \cite{GSW04}. 

How important are the different one-loop contributions? For $\xi\gsim1$
the cubic term (\ref{cubic}) is still relevant: Leaving it out considerably 
weakens the phase transition from $\xi=1$ to $\xi=0.56$,  for $m_H=115$ GeV.
Omitting also the log-term (\ref{log}) makes the phase
transition stronger again, $\xi=0.81$. In addition, getting rid of the
one-loop term of eq.~(\ref{1_2loop}) increases $\xi$ to 1.27.
Thus the one-loop contributions in (\ref{cubic}) - (\ref{1_2loop}) partially
cancel each other and therefore our results agree reasonably well with those of
ref.~\cite{GSW04}. For larger Higgs masses and stronger phase transition the picture 
is qualitatively similar, however, the cubic term becomes less important.

The two-loop $ \phi  ^ 2 $ term of eq.~(\ref{1_2loop}) practically does not change the
result, demonstrating that the dimension-6 operator does not spoil
the loop expansion. We have also checked that adding a dimension-8 term
$(1/M^4)(\Phi^{\dagger}\Phi)^4$ only affects $\xi$ at the order of $v^2/M^2$.     

The one-loop effective potential was also discussed in ref.~\cite{HO04}.
However, the authors impose an erroneous lower bound on the
cut-off scale, requiring a positive mass squared for the
Goldstone boson. As a result, they obtain a lower bound on the
Higgs mass from eq.~(\ref{wo}) which is much too small.


%
%
\section{The bubble properties}
In this section we discuss some bubble properties which 
will enter the computation of the baryon asymmetry, in particular
the  thickness $L_{\rm w}$, and the velocity, $v_{\rm w}$ of the wall.
As already mentioned, the two minima of $V_{\rm eff}$ become of the same 
depth at $T_c$, but tunneling 
with the formation of bubbles of the broken phase
will start somewhat later at a temperature $T_n$. The probability for
thermal tunneling is exponentially suppressed by the energy of the critical bubble, $S_3$.
The phase transition starts if the nucleation probability per horizon volume
becomes of order unity, which translates to $S_3(T_n)/T_n\sim130-140$ \cite{AH92}.

For $\xi=1$ the amount of supercooling, i.e. the difference between the critical and nucleation
temperatures,  is small. For $m_H=115$ GeV we find $T_c=107.34$ GeV and 
$T_c-T_n=0.45$ GeV.
The system is well described by the thin wall approximation.
For smaller values of $M$ and stronger phase transition, supercooling becomes
more and more important. The thin wall approximation is no longer reliable.
At some critical value the phase transition does no longer proceed at all.
The universe remains stuck in the symmetric vacuum. This regime is indicated
by the "metastability" line in fig.~\ref{xi-lines}.

\begin{figure}[t]
\begin{center}
   \epsfig{file=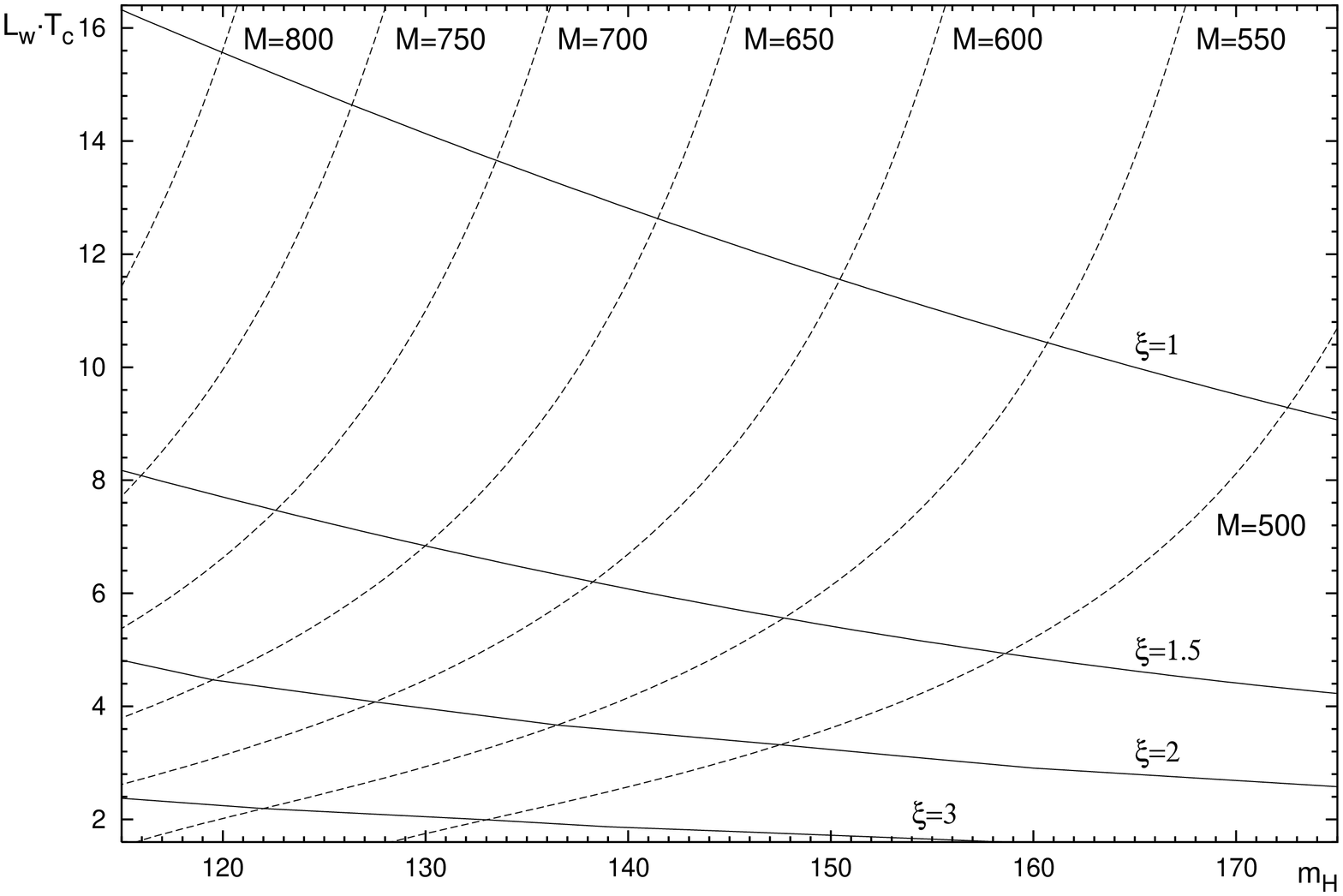,width=80mm,angle=270}
\end{center}
\caption{The wall thickness $L_{\rm w}$ as a function of the Higgs-mass $m_H$ for
         several values of $M$, which are given in units of GeV. In addition lines of
         constant $\xi$ are shown.}
\label{LW}
\end{figure}

Once a critical bubble is nucleated it will expand. The expansion is accelerated
by the internal pressure and slowed down by plasma friction. Finally, a
stationary situation will be reached, where the different forces are balanced, and 
the wall propagates with constant velocity, $v_{\rm w}$. In order to estimate the 
thickness of the bubble wall, we ignore friction for a moment and solve the field
equation at the critical temperature with the effective potential of eq.~(\ref{Veff}),
\be \label{eom2}
{d^2\phi\over dz^2}={\partial \over \partial
  \phi}V_{\rm eff}(\phi).
\ee
The boundary conditions read $ \phi(z \rightarrow -\infty)=v_c$ and 
$\phi(z \rightarrow \infty)=0$.
The bubble profile can approximately be described
by a kink,
\be
\phi(z)={v_c\over 2}\left(1-\mbox{tanh}{z\over L_{\rm w}}\right)
\ee
with $L_{\rm w}=\sqrt{v_c^2/(8V_b)}$, where $V_b$ is the height of the potential
barrier. This relation would be exact for a $\phi^4$ potential and 
we found that it is also a good approximation in our case.
In fig.~\ref{LW} the wall thickness is shown in dependence of the Higgs 
mass $m_H$ and $M$. As we decrease $M$ at fixed $m_H$, the wall thickness 
in units of $1/T_c$ becomes smaller. The same happens if we decrease
$m_H$ at fixed $M$. The main effect comes from the decrease of $T_c$ 
in these cases. Notice that $L_{\rm w}T_c$ varies in a wide range
between about 2 and 16.  

Let us finally comment on the wall velocity. Taking into account only the 
friction related to the infrared gauge field modes \cite{M00}, 
\begin{equation}\label{v_w}
v_{\rm w}=\frac{32\pi L_{\rm w}}{11g_2^2T^3}\cdot\frac{\Delta V}{\ln(m_WL_{\rm w})+{\cal O}(1) }
\end{equation}
we obtain a wall velocity of order unity, except for $\xi$ being close to one.
Here $\Delta V$ is the potential difference
at the nucleation temperature and $m_W$ the mass of the W boson. The order unity
correction in the denominator
is induced by friction of other particles in the plasma, in particular the top quark
\cite{MP95}. Since numerically $\ln(m_WL_{\rm w})$ is only of order unity, top quark
friction will slow down the wall considerably. The wall velocity is
further reduced by latent heat of the nucleating bubbles. In general, the
wall moves faster in the case of a stronger phase transition.

Let us briefly discuss two representative examples. Taking $\xi=1$ and
$m_H=115$ GeV, we obtain $v_{\rm w}=0.24$ from eq.~(\ref{v_w}). Including
the effect of reheating, $\Delta V$ is reduced and we arrive at
$v_{\rm w}=0.08$. If we finally switch on the order unity correction,
we end up with $v_{\rm w}\sim0.05$. The picture looks very much different
for stronger phase transitions. Going to $\xi=1.5$, eq.~(\ref{v_w}) already
leads to a wall velocity of order unity. Including again top quark
friction and reheating, we obtain $v_{\rm w}\sim0.5$. For larger Higgs masses
we find a very similar behavior. These are only very rough estimates,
since  eq.~(\ref{v_w}) breaks down for large values of $v_{\rm w}$.
Given these uncertainties we will treat $v_{\rm w}$ as a free parameter in our
computation of the baryon asymmetry.


%
%
\section{CP violation}
Non-renormalizable interactions provide new sources of CP violation. 
In the absence of gauge singlets the leading operators are of mass dimension six.
In ref.~\cite{DHSS91} a $|\Phi|^2F\tilde F$ operator was discussed. We
will focus on the operators 
\begin{equation} \label{xij}
\frac{x_{ij}^u}{M^2}(\Phi^{\dagger}\Phi)\Phi u^c_iq_j
\end{equation}
which have been proposed to drive baryon number generation in ref.~\cite{ZLWY94}. 
There are analogous terms for the down-type quarks and leptons. 
The fermion masses become
\begin{equation}  \label{mij}
m_{ij}=y_{ij}\frac{v}{\sqrt{2}}+\frac{v^3}{2\sqrt{2}M^2}x_{ij}
\end{equation}
where $y_{ij}$ are the ordinary Yukawa couplings. In unitary gauge,
the effective Yukawa couplings to the physical Higgs boson are given by
\begin{equation}  \label{Yij}
Y_{ij}=y_{ij}+\frac{3v^2}{2M^2}x_{ij}.
\end{equation}
Thus, there is a mismatch of order $(xv^2/M^2)$ between the
fermion masses and the effective Yukawa couplings.
In general, the 
couplings $x_{ij}$ contain complex phases, and they are of unknown
flavor structure.  While for the top quark $x_{33}^u$ may 
be of order unity, the couplings of the lighter fermions should not exceed
${\cal O}(M^2m_f/v^3)$ to avoid fine tuning of the fermion masses $m_f$. 
For instance, the corresponding coupling for the electron should at most be 
of order $10^{-4}\times(M/\rm{TeV})^2$, which is a small number. 
Having this in mind, and lacking a theory of flavor mixing, we will therefore 
assume that the $x_{ij}$ have a similar flavor structure as the corresponding
Yukawa couplings, i.e.~$x_{ij}\sim y_{ij}$, up to order unity coefficients.
This structure could be motivated by a Froggatt-Nielsen \cite{froggatt} 
type mechanism, where
the operators (\ref{xij}) and the ordinary Yukawa couplings would have 
the same quantum numbers.

Since the operators (\ref{xij}) do not have to be strictly aligned with the
ordinary Yukawa couplings, they will induce flavor violation and CP violation.
Via tree-level Higgs exchange they generate four-fermion interactions, which, for instance,
affect $K-\bar K$ mixing. For example, the operator $(1/Q^2)(d^cs\bar d \bar{s^c})$ is 
generated at a level of
\begin{equation} \label{ds}
\frac{1}{Q^2}\sim\frac{v^4}{M^4m_H^2}x_{12}^d(x_{21}^d)^*.
\end{equation}
For $x_{12}^d\sim m_s\sin\theta_c/v$,
where $m_s$ denotes the strange quark mass and $\theta_c$ the
Cabibbo angle, we obtain roughly $Q\sim 5\cdot10^7$GeV$\times(M/\rm{TeV})^2$.
At $M\gsim30$ TeV a 2-loop contribution of order 
$(1/(16\pi^2)^2)x_{12}^d(x_{21}^d)^*/M^2$ becomes dominant.\footnote{This 
diagram is quadratically
divergent which we cut off at the scale $M$.} Experimentally, these operators
are constrained to be suppressed by $Q\gsim10^7$ GeV, especially in the
presence of CP violation. This constraint is compatible with (\ref{ds}) for
$M\gsim500$ GeV.
Later on
we will only be interested in the coupling of the top, $x_{33}^u\sim1$. If we make the
unnatural assumption that only this coupling is present, a $dd\bar s\bar s$
operator is generated at 2-loops with $Q\sim 4\cdot10^{9}$GeV$\times(M/\rm{TeV})^2$.
In the $B$ system the experimental bounds are even more easily satisfied.

CP violating couplings induce electric dipole moments (EDM's). The EDM
of the neutron, for instance, is experimentally constrained by 
$d_n/e<6.3\cdot10^{-26}$ cm \cite{d_n}.
The individual EDM's of up and down quarks should therefore be not much
larger. The up quark receives the larger contribution. At one-loop we find
\begin{equation}
\frac{d_u}{e}\sim \frac{1}{16\pi^2}\frac{v^3
{\rm Im}(x_{13}^ux_{31}^u)}{M^4}\sim1\cdot10^{-26} {\rm cm}
\times\left(\frac{\rm{TeV}}{M}\right)^2,
\end{equation} 
which might be close to the experimental bound. In the second step we assumed
a maximal phase and $|x_{13}^u|\sim|x_{31}^u|\sim  V_{ub}$.
If only the coupling $x_{33}^u$ is present, an EDM of roughly 
$d_u/e\sim 5\cdot10^{-27} {\rm cm}\times {\rm Im}(x_{33}^u)\times(\rm{TeV}/M)^2$ 
is induced at the 2-loop level \cite{ZLWY94}.
Thus a coupling $x_{33}^u$ of order unity with a large phase can be present
without inducing too large flavor changing neutral currents or EDM's, even
for $M\sim300$ GeV.  Other
couplings also are allowed, as long as some Yukawa-like 
hierarchy is respected. However, experimental signals of non-standard flavor
physics could be detected in the near future.

At low energies the non-renormalizable operators and the Yukawa couplings
melt into the couplings (\ref{Yij}). 
However, during the EWPT 
the two terms in (\ref{Yij}) vary differently along the bubble wall. As
a result, the fermion masses acquire position dependent phases which
cannot be rotated away. For the phase of the top quark mass we obtain 
\begin{equation} \label{theta}
\tan\theta_t(z)\approx \sin\varphi_t\frac{\phi^2(z)}{2M^2}\frac{x_t}{y_t},
\end{equation} 
where we defined $x_te^{i\varphi_t}\equiv x_{33}^u$ and ignored the
real part of $x_{33}^u$. In two Higgs doublet models such a phase may
arise from spontaneous CP violation. In supersymmetric models position 
dependent phases are induced by flavor mixing, e.g.~for the charginos. 
In the next section we discuss how the phase (\ref{theta}) drives the
generation of a baryon asymmetry as a bubble wall moves through the
plasma. 

%
%
\section{Transport equations}
The CP violating interactions of particles in the plasma with the bubble
wall create an excess of left-handed quarks over the corresponding
anti-quarks.\footnote{Quarks will turn out to be more important than
leptons because of the large top mass.} 
In the symmetric phase the left-handed quark density biases the
sphaleron transitions to generate a net baryon asymmetry.

In the WKB approach the CP violating interaction of a fermion 
with the wall leads to different dispersion relations for
particles and anti-particles \cite{JPT95}, depending on their
complex masses. To make this method applicable, it is required that the
typical de Broglie wavelength of particles in the plasma is small
compared to the width of the wall, i.e.~$L_{\rm w}T\gg1$. Otherwise
an expansion in derivatives of the background Higgs field cannot be justified.
According to the results of section 3 this is a good approximation
in a large fraction of our parameter space. It is violated only
in the cases of a very strong phase transition, $\xi\gsim3$. 
From the dispersion relations we compute a force
term which enters the transport equations that describe
the evolution of the plasma. An alternative approach was followed
in ref.~\cite{CQW}.
 
Let us consider a single Dirac fermion, such as the top quark,  
with a space-time dependent mass 
${\rm Re}{\cal M}(\bar z)+i\gamma^5 {\rm Im}{\cal M}(\bar z)$, 
where ${\cal M}(\bar z)=m(\bar z)e^{i\theta(\bar z)}$ and $\bar z\equiv z-v_{\rm w}t$ 
denotes the relative coordinate perpendicular to the wall. For particles and anti-particles 
the dispersion relations to first order in derivatives read \cite{CJK00}
\begin{equation} \label{disp}
E_{\pm}=E_0\pm\Delta E=\sqrt{p^2+m^2}\pm {\rm sign}(p_z)\theta'
\frac{m^2}{2(p^2+m^2)},
\end{equation}
where $p^2=\vec{p}^2$ is the squared kinetic momentum and
$\theta'= d \theta/d \bar z$. $E_+$ is the
energy of left-handed particles, $E_-$ corresponds to
the right-handed states, and for the anti-particles the other way round.\footnote{Later on the relevant particles will be relativistic,
so that we can approximate helicity by chirality.} 
In a more rigorous treatment similar dispersion relations 
were derived for spin states in the Schwinger-Keldysh 
formalism \cite{PSW,PSW2}. 

The evolution of the particle distributions $f_i(t,\vec{x},\vec{p})$ 
we describe by classical Boltzmann equations. 
The dispersion relations 
(\ref{disp}) induce force terms, which are different for particles and 
anti-particles.
To make the system of equations tractable, we use a fluid-type
ansatz in the rest frame of the plasma \cite{JPT95}
\begin{equation}
  f_i(t,\vec{x},\vec{p})=\frac{1}{e^{\beta(E_i-v_ip_z-\mu_i)}\pm1},
\end{equation} 
where $v_i$ and $\mu_i$ denote velocity perturbations and 
chemical potentials for each fluid. The velocity perturbations
are introduced to model the movement of particles in response to
the force. 

For a shorter notation let us first introduce some symbols $K$, which represent
momentum averages normalized relative to the massless Fermi-Dirac case, 
\begin{equation}
\langle X \rangle \equiv \frac{\int d^3p X(p)}{\int d^3p f_+'(m=0)},
\end{equation}
where $f_{\pm}'=-\beta e^{\beta E_0}/(e^{\beta E_0}\pm1)^2$. We define
\bqa
&&K_{1,i}=\left<{p_z^2\over \sqrt{p^2+m_i^2}} f_{\pm}'(m_i)\right>, \quad
K_{2,i}=\left<p_z^2 f_{\pm}'(m_i)\right>,\nn\\[.2cm]
&&K_{3,i}=\left<{1\over 2\sqrt{p^2+m_i^2}} f_{\pm}'(m_i)\right>,\quad
K_{4,i}=\left<{|p_z|\over 2(p^2+m_i^2)} f_{\pm}'(m_i)\right>,\nn\\[.2cm]
&&K_{5,i}=\left<{|p_z|p^2\over 2(p^2+m_i^2)^2} f_{\pm}'(m_i)\right>,\quad
K_{6,i}=\left<\left({|p_z|\over (p^2+m_i^2)^2}-{\delta(p_z)\over
    p^2+m_i^2}\right) f_{\pm}'(m_i)\right>,\nn\\[.2cm]
&&K_{7,i}=\left<{|p_z|^3\over (p^2+m_i^2)^2} f_{\pm}'(m_i)\right>,
\eqa
which appear in the transport equations discussed in the following.
In the case of a massless fermion we obtain $K_1=1.096T$, $K_2=4.606T^2$, 
$K_3=0.211/T$, $K_4=0.105/T$, $K_5=0.105/T$, $K_6=-0.038/T^3$ and $K_7=0.105/T$.
For $m\gg T$ the averages experience an exponential Boltzmann
suppression.

We look for solutions of the Boltzmann equation which are stationary,
i.e.~which only depend on the relative coordinate $\bar z$. We
expand the Boltzmann equation in derivatives of the fermion mass.
At first order in derivatives there is no difference between particles
and anti-particles. Weighting the Boltzmann equation with 1 and $p_z$, we
obtain after momentum averaging 
\begin{eqnarray} \label{1a}
\kappa_iv_{\rm w}\mu_{i,1}'-K_{1,i} v_{i,1}'-\langle {\cal C}_i\rangle
&=&K_{3,i} v_{\rm w} (m_i^2)'
\\[.3cm] \label{1b}
-K_{1,i} \mu_{i,1}'+K_{2,i} v_{\rm w} v_{i,1}'-\langle p_z{\cal C}_i\rangle&=&0.
\end{eqnarray}
Here $\mu_{i,1}$ and  $v_{i,1}$ indicate the perturbations to first order 
in derivatives. A prime denotes again a derivative with respect to $\bar z$.
The statistical factor 
$\kappa_i\equiv \langle f_{\pm}'(m_i) \rangle$ is 1 (2) for massless
fermions (bosons) and becomes exponentially small for $m\gg T$. 
The force term on the right-hand side is induced by the change in the
particle mass along the wall.
Introducing
inelastic rates, $\Gamma_p^{\rm inel}$, and elastic rates, $\Gamma^{\rm el}_p$, for a process $p$,
the collision terms take the form \cite{CJK98}
\begin{equation}
\langle {\cal C}_i \rangle=\sum_p \Gamma_p^{\rm inel}\sum_j\mu_{j}, 
\quad 
\langle p_z{\cal C}_i\rangle=v_{i}\bar p_z^2\sum_p \Gamma^{\rm el}_p\equiv v_{i}\bar p_z^2 \Gamma^{\rm el}_i.
\end{equation}
In $\langle p_z{\cal C}_i\rangle$ we neglected inelastic processes. 
The leading order eqs.~(\ref{1a}), (\ref{1b}) contain the friction terms which enter 
the computation of the wall velocity \cite{MP95}.

To second order in derivatives, we have to distinguish between particles 
and anti-particles. The perturbations contain CP odd and even components,
\begin{equation}
\mu_i=\mu_{i,1}+\mu_{i,2o}+\mu_{i,2e},\quad 
v_i=v_{i,1}+v_{i,2o}+v_{i,2e}.
\end{equation}
In the following only the odd second order perturbations will enter, so we can drop the 
subscript ``$o$'' to simplify the notation.
Subtracting the equations of particles and anti-particles, we obtain
\begin{eqnarray} \label{2a}
\kappa_iv_{\rm w}\mu_{i,2}'-K_{1,i} v_{i,2}'-\langle {\cal C}_i\rangle
=-K_{6,i} \theta_i'm_i^2\mu_{i,1}'~~~~~~~~~~~~~~~~~~~~~~~~~~~~~~~~~~~~~~~~
\\[.3cm] 
-K_{1,i} \mu_{i,2}'
+K_{2,i} v_{\rm w} v_{i,2}'-\langle p_z{\cal C}_i\rangle
= K_{4,i} v_{\rm w}m_i^2\theta_i'' 
+K_{5,i} v_{\rm w}(m_i^2)'\theta_i'
-K_{7,i} m_i^2\theta_i'v_{i,1}'.~~
\label{2b}
\end{eqnarray}
Note that the CP violating source terms are proportional to derivatives of $\theta_i$.
A constant phase does not contribute.
The source terms proportional to the first order perturbations have not been
investigated so far in a realistic context. (See ref. \cite{PSW2} for a discussion in
the context of Schwinger-Keldysh formalism.) To study their
relevance for the generation of the observed baryon asymmetry will be
a main issue in the next section. 

We may use eq.~(\ref{2a}) to solve for $v_{i,2}$. Neglecting derivatives
of the thermal averages, which are higher order in derivatives,
we end up with diffusion equations for the chemical potentials
\begin{eqnarray} 
-\kappa_iD_i(1-A_iv_{\rm w}^2)\mu_{i,2}'' -\kappa_iv_{\rm w}\mu_{i,2}'-
D_iA_iv_{\rm w}\sum_p \Gamma_p^{\rm inel}\sum_j\mu_{j,2}' &&
\nonumber\\ \label{dif}
+\sum_p \Gamma_p^{\rm inel}\sum_j\mu_{j,2}
-A_iD_iv_{\rm w}\sum_p (\Gamma_p^{\rm inel})'\sum_j\mu_{j,2}&=&S_i,
\end{eqnarray}
where 
\begin{equation}
A_i=\frac{\kappa_i K_{2,i}}{K_{1,i}^2}.
\end{equation}
In the massless limit we have $A\approx 3.83$. The diffusion constants
are given by  \cite{CJK98}
\begin{equation}
\kappa_i D_i=\frac{K_{1,i}^2}{\bar p_z^2\Gamma^{\rm el}_i}.
\end{equation}
To leading order in the wall velocity, neglecting derivatives of the inelastic rates
and ratios of inelastic to elastic rates, the left-hand side of eq.~(\ref{dif}) 
reproduces the result of ref.~\cite{CJK98}. This corresponds to dropping the
terms proportional to $A_i$.
In the next section we will examine
to what extent this simplification is justified. Since $A_i$ is not a small number,
the corrections will turn out to be important in certain regimes. 
For the source terms we obtain
\begin{eqnarray} \label{source}
S_i&=&\frac{\kappa_iD_iv_{\rm w}}{K_{1,i}}
\left(
K_{4,i} m_i^2\theta_i''  
+K_{5,i}(m_i^2)'\theta_i'
\right)'\nn\\
&&+K_{6,i}\left(1-AD_iv_{\rm w}\frac{d}{d\bar z}\right)
\left(m_i^2\theta_i'\mu_{i,1}'\right)
-\frac{\kappa_iD_iK_{7,i}}{K_{1,i}}
\left(m_i^2\theta_i'v_{i,1}' \right)'.
\end{eqnarray}
In the second line new source terms related to the first order
perturbations are showing up. As expected, no source terms are left in the
case of vanishing wall velocity. 

Let us apply these general results to the SM with a low cut-off. In a 
first step we compute the asymmetry in the left-handed quark density. 
At this stage we neglect the weak sphalerons, i.e.~baryon 
and lepton number are conserved. The most important particle species are the left- and 
right-handed top quarks and the Higgs bosons. Leptons are only
produced by small Yukawa couplings and therefore not
taken into account. It turns out that also the Higgs bosons
have only a minor impact on the generated baryon asymmetry.
They change the final result only at the percent level, so we can
ignore them.
In a second step, the weak sphalerons convert
the left--handed quark number into a baryon asymmetry.

We take into account the top Yukawa interaction, $\Gamma_y$, the strong sphalerons,
$\Gamma_{ss}$ and the top helicity flips, $\Gamma_m$ caused by the interactions
with the bubble wall.
The latter are only present in the broken phase.
The gauge interactions are assumed to be in equilibrium. The transport
equations become
\begin{eqnarray}\label{eq1}
(3\kappa_t+3\kappa_b)v_{\rm w}\mu_{q_3,2}'-(3K_{1,t}+3K_{1,b})v_{q_3,2}'
-6\Gamma_y\left(\mu_{q_3,2}+\mu_{t,2}\right)
\nn\\
-6\Gamma_m\left(\mu_{q_3,2}+\mu_{t,2}\right)
-6\Gamma_{ss}\left[\left(2+9\kappa_t+9\kappa_b\right)
\mu_{q_3,2}+\left(1-9\kappa_t\right)\mu_{t,2}\right]
\nn\\
=-3K_{6,t}m^2_t\theta_t'\mu_{q_3,1}'
\end{eqnarray}
\begin{eqnarray}
-(K_{1,t}+K_{1,b})\mu_{q_3,2}'+(K_{2,t}+K_{2,b})v_{\rm w}v_{q_3,2}'-
\left({K_{1,t}^2\over \kappa_tD_{Q}}+{K_{1,b}^2\over
  \kappa_bD_{Q}}\right)v_{q_3,2}
\nn\\
=K_{4,t}v_{\rm w}m^2_t\theta_t'' 
+K_{5,t}v_{\rm w}(m^2_t)'\theta_t'-K_{7,t}m^2_t\theta_t'v_{q_3,1}' 
\end{eqnarray}
\begin{eqnarray}
3\kappa_tv_{\rm w}\mu_{t,2}'-3K_{1,t}v_{t,2}'-6\Gamma_y\left(\mu_{q_3,2}+\mu_{t,2}\right)
-6\Gamma_m\left(\mu_{q_3,2}+\mu_{t,2}\right)
\nn\\
-3\Gamma_{ss}\left[\left(2+9\kappa_t+9\kappa_b\right)
\mu_{q_3,2}+\left(1-9\kappa_t\right)\mu_{t,2}\right]
\nn\\
=-3K_{6,t}m^2_t\theta_t'\mu_{t,1}'
\end{eqnarray}
\begin{eqnarray}
-K_{1,t}\mu_{t,2}'+K_{2,t}v_{\rm w}v_{t,2}'-{K_{1,t}^2\over
  \kappa_tD_Q}v_{t,2}
\nn\\ \label{eq4}
=K_{4,t}v_{\rm w}m^2_t\theta_t'' +K_{5,t}v_{\rm w}(m^2_t)'\theta_t'-
K_{7,t}m^2_t\theta_t'v_{t,1}'.
\end{eqnarray}
The top quark phase, $\theta_t$, is given by eq.~(\ref{theta}).
For the chemical potentials we take  
$\mu_t=\mu(u^c_3)$ and $\mu_{q_3}=(\mu(u_3)+\mu(d_3))/2$.
The index $t$ ($b$) refers to the top
and bottom quark, respectively.
We have omitted the tiny source of the bottom quark which is suppressed
by $(m_b/m_t)^4$. We used baryon number conservation to express
the strong sphaleron interaction in terms of $\mu_{q_3,2}$ and $\mu_{t,2}$
\cite{HN95}. Replacing the second order source terms by the first order ones,
the same system of transport equation holds for the first order perturbations.

Using again baryon number conservation, the chemical potential of 
left-handed quarks, $\mu_{B_L}=\mu_{q_1}+ \mu_{q_2}+\mu_{q_3}$,
is obtained as
\begin{equation}
\mu_{B_L}=(1+2\kappa_t+2\kappa_b)\mu_{q_3}-2\kappa_t\mu_t.
\end{equation}
The baryon asymmetry is then given by \cite{CJK00}
\begin{equation} \label{eta1}
\eta_B=\frac{n_B}{s}=\frac{405\bar\Gamma_{ws}}{4\pi^2v_{\rm w}g_*T^4}\int_0^{\infty}
d\bar z \mu_{B_L}(\bar z)e^{-\nu\bar z},
\end{equation}
where $\bar\Gamma_{ws}$ is the weak sphaleron rate and
$\nu=45\bar\Gamma_{ws}/(4v_{\rm w}T^3)$. The effective number of
degrees of freedom in the plasma is $g_*=106.75$.
In eq.~(\ref{eta1}) the
weak sphaleron rate has been suddenly switched off in the
broken phase, $\bar z<0$. The exponential factor in the integrand
accounts for the relaxation of the baryon number if the wall
moves very slowly.

%
%
\section{Numerical Results}
In this section we present our evaluations of the transport equations 
(\ref{eq1}) - (\ref{eq4}). We will discuss under what conditions the
terms proportional to $A_i$ may be neglected in eq.~(\ref{dif}) and investigate 
what is the impact of the new source terms in  eqs.~(\ref{2a}) and (\ref{2b}).
Finally, we will demonstrate that the SM with a low cut-off can explain
the observed baryon asymmetry for natural values of the parameters.

In our numerical computations we use the following values for the 
weak sphaleron rate \cite{Mws}, the strong sphaleron rate \cite{Mss}, the top
Yukawa rate \cite{HN95}, the top helicity flip rate and the quark diffusion
constant  \cite{HN95}
\begin{eqnarray}
&&\bar \Gamma_{ws}=1.0\cdot10^{-6}T^4, \quad  \quad\bar \Gamma_{ss}=4.9\cdot10^{-4}T^4, 
\nn \\
&&\Gamma_y=4.2\cdot10^{-3}T, \quad \quad ~~~\Gamma_m=\frac{m_t^2(\bar z,T)}{63T},
\nn \\
&&D_Q=\frac{6}{T}.
\end{eqnarray}
Note that in $\bar \Gamma_{ss}\equiv \Gamma_{ss}T^3$ the value 
$\alpha_s=0.086$ from the dimensionally reduced theory has been used \cite{Mss}. 
Changing the rates $\Gamma_y$ and $\Gamma_m$ has only a small
effect on the baryon asymmetry, as would have the inclusion of the Higgs
field chemical potential in the transport equations. Doubling the value of
$D_Q$ enhances the baryon asymmetry by 20-30\% because of more
efficient diffusion. Enhancing $\Gamma_{ss}$ reduces the baryon asymmetry
since the strong sphalerons drive $\mu_{B_L}$ to zero if the quarks
are taken massless \cite{GS94}. The baryon asymmetry also depends on whether
we take the top quark to be massive or massless in the thermal averages.
If we switch on the top mass, the baryon asymmetry becomes smaller
since the thermal averages go down. In the evaluations we use the half
$m_t^2$ of the broken phase to compute the averages.  

\FIGURE[t]{
   \epsfig{file=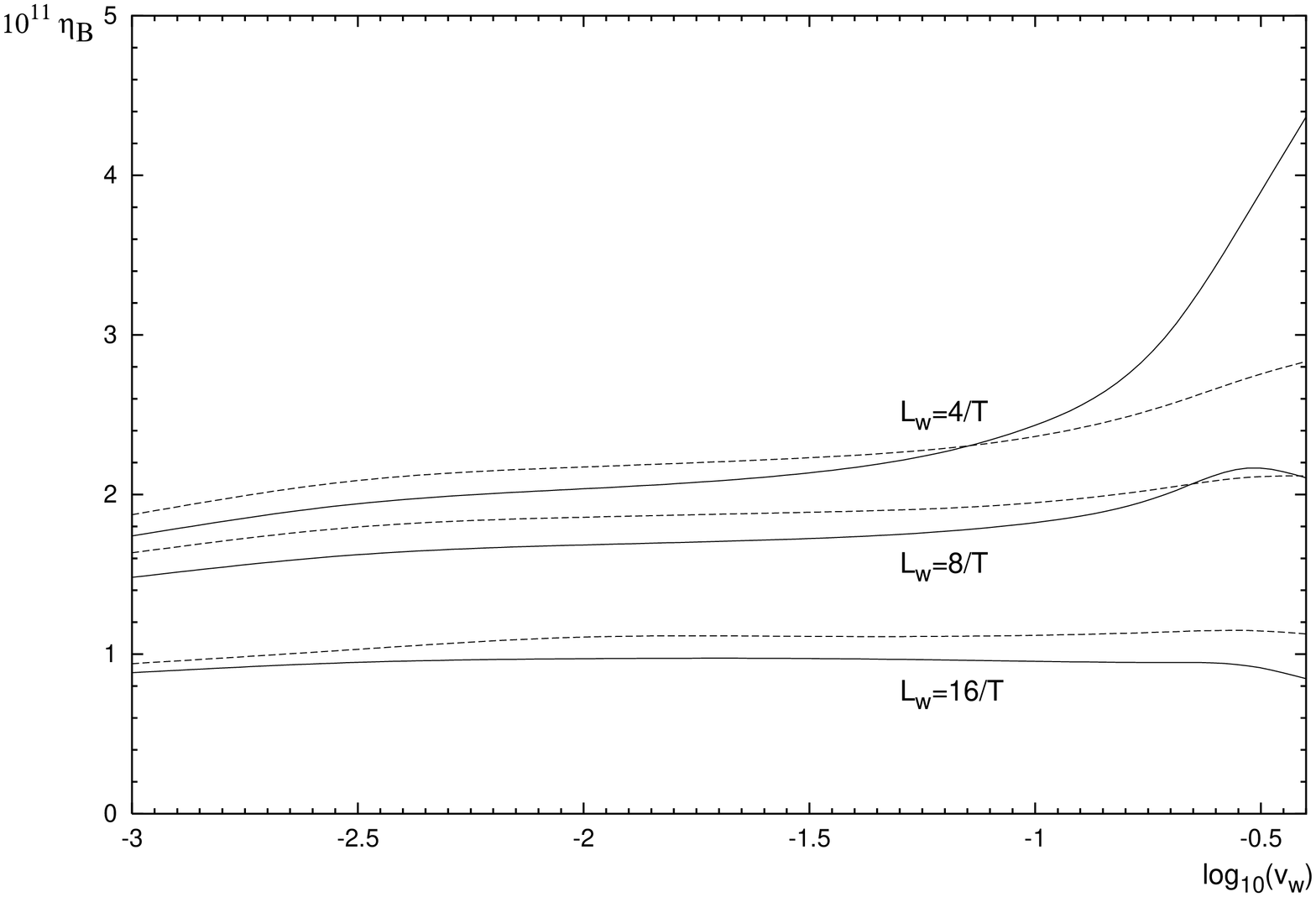,width=80mm,angle=270}
\caption{Comparison between eq.~(\ref{dif}), where the terms proportional to $A_i$
have been neglected (dashed), and eqs.~(\ref{eq1}) - (\ref{eq4}) without the
first order perturbations (solid) for different values of $L_{\rm w}$. The other parameters are taken as $\xi=1.5$ and $M=6T$.
}
\label{3_6}
}
Let us first discuss under which conditions the $A_i$ corrections in  eq.~(\ref{dif})
become important. At this stage we do not yet relate the bubble wall parameters
to the model introduced in section 2.
In fig.~\ref{3_6} we display the baryon asymmetry
computed with the simplified equations (dashed lines) compared to the one obtained 
from the full equations (\ref{eq1}) - (\ref{eq4}) (solid lines) as a function of the wall velocity. We take $\xi=1.5$ and $M/T=6$ 
for three different values of $L_{\rm w}$. The CP violating
phase in the dimension-6 operator (\ref{xij}) we take as maximal,
i.e.~$\sin\varphi_t=1$, and we choose $x_t=1$. In any case the baryon asymmetry
is proportional to $x_t\sin\varphi_t$. 
Since we want to compare the left hand sides of the transport
equations, we included only the source term of the first line of eq.~(\ref{source}).
The simplified equations give a reasonable description for $v_{\rm w}\lsim0.1$. 
For large values of $v_{\rm w}$ there are sizable deviations, especially
for thinner bubble walls. This behavior is expected since the $A_i$ corrections come with
additional powers of the wall velocity. In the MSSM, where 
$v_{\rm w}\sim0.05-0.1$ \cite{john},
the simplified equations are applicable. In the following we will use the
full equations (\ref{eq1}) - (\ref{eq4}) to compute the baryon asymmetry. 
Fig.~\ref{3_6} also demonstrates that the baryon asymmetry increases for thinner
bubble walls. This behavior is expected since the source terms involve
derivatives of the background Higgs field. 

In fig.~\ref{old_new_v} we compare the contributions to the baryon asymmetry due
to the different source terms on the right hand side of eqs. (\ref{2a}) and (\ref{2b}), using the parameters of fig.~\ref{3_6} and $L_{\rm w}=8/T$. 
The new source terms proportional to the first order perturbations $\mu_{i,1}$ and $v_{i,1}$ 
are non-negligible. They enhance the baryon asymmetry, especially for
small values of $v_{\rm w}$. For large wall velocities they do no longer matter.
The total baryon asymmetry depends only mildly on $v_{\rm w}$, which is quite positive,
given our poor understanding of this parameter. For other wall widths the 
picture is similar.
 
As shown in fig.~\ref{old_new_xi} the baryon asymmetry increases rapidly for
larger values of $\xi$. We fixed $v_{\rm w}=0.01$ and 0.3 and again $L_{\rm
  w}=8/T$. For large $\xi$ the 
top quark mass becomes larger in the broken phase,
and the source terms involve powers of $m_t^2$. Also the CP violating phase in the
top quark mass from eq.~(\ref{theta}) becomes stronger. The source terms related to the
first order perturbations have an additional power of $m_t^2$. Therefore they grow
even faster and dominate for large $\xi$. This behavior holds also for other
values of $L_{\rm w}$.

\FIGURE[t]{
   \epsfig{file=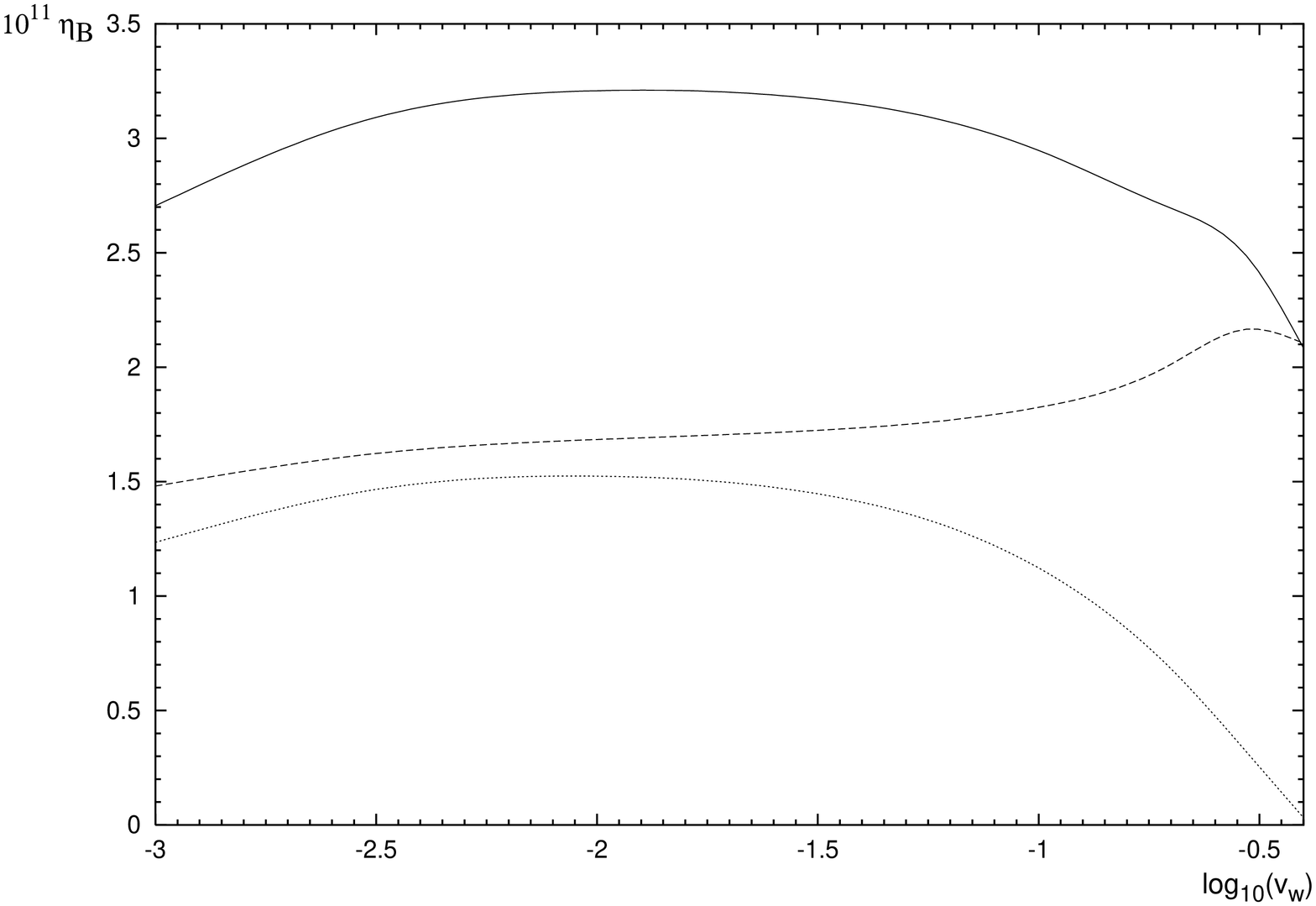,width=80mm,angle=270}
\caption{The solid line
  represents $\eta_B$ as a function of the wall velocity for $L_{\rm w}=8/T$,
  $M=6T$ and $\xi=1.5$. The dashed line would be the asymmetry without the source
  terms containing the first order perturbations $\mu_1'$ and $v_1'$ 
  and the dotted one is the contribution due to these terms only.
}
\label{old_new_v}
}
\FIGURE[t]{
   \epsfig{file=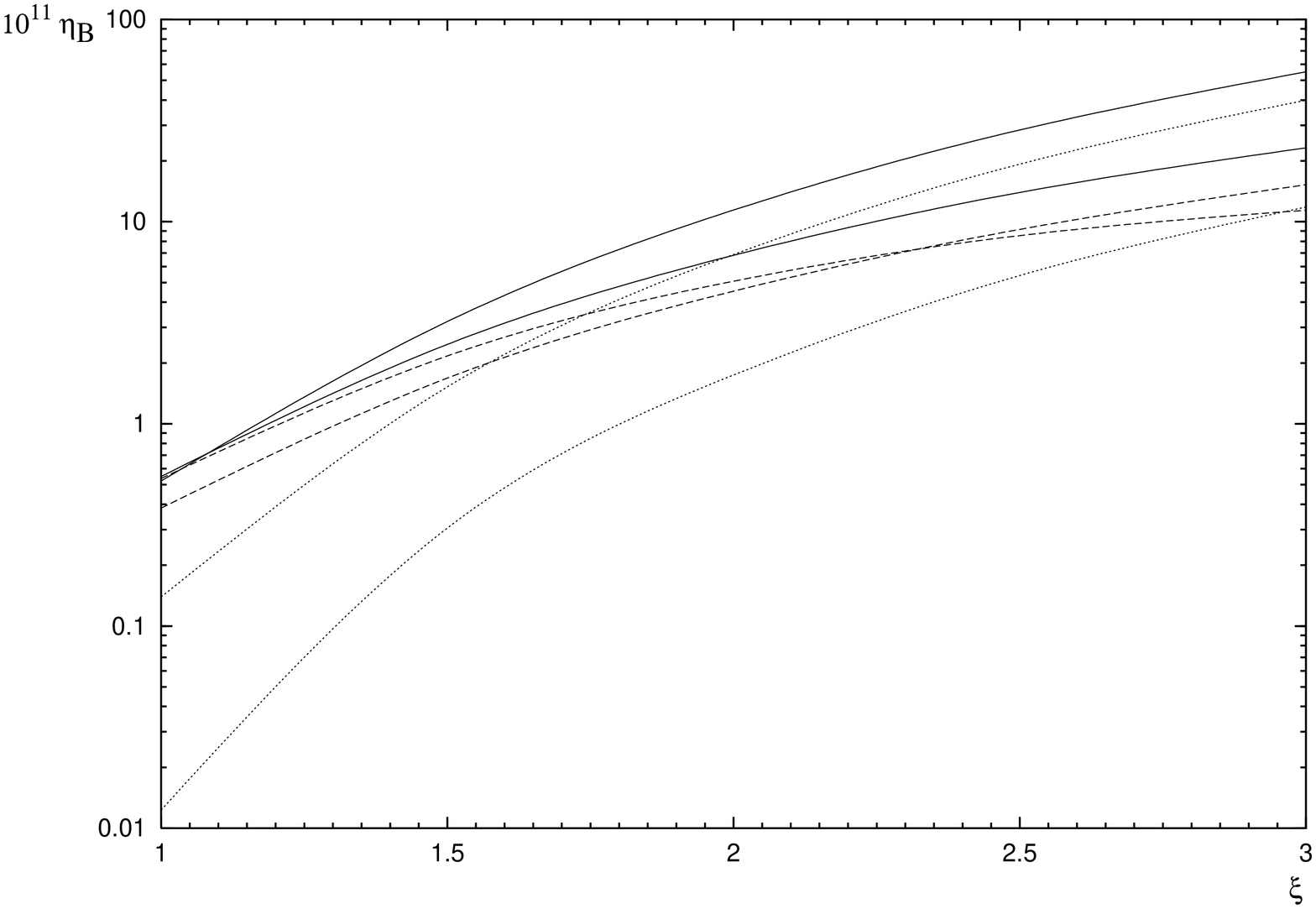,width=80mm,angle=270}
\caption{
The baryon asymmetry as a function of $\xi$ for two wall velocities. The
  different line types have the
  same meaning as in fig.~\ref{old_new_v}, and again $L_{\rm
  w}=8/T$ and $M=6T$. On the right edge of the figure the upper curves are for
  $v_{\rm w}=0.01$ and the lower ones for $v_{\rm w}=0.3$.}
\label{old_new_xi}
}
\FIGURE[h]{
   \epsfig{file=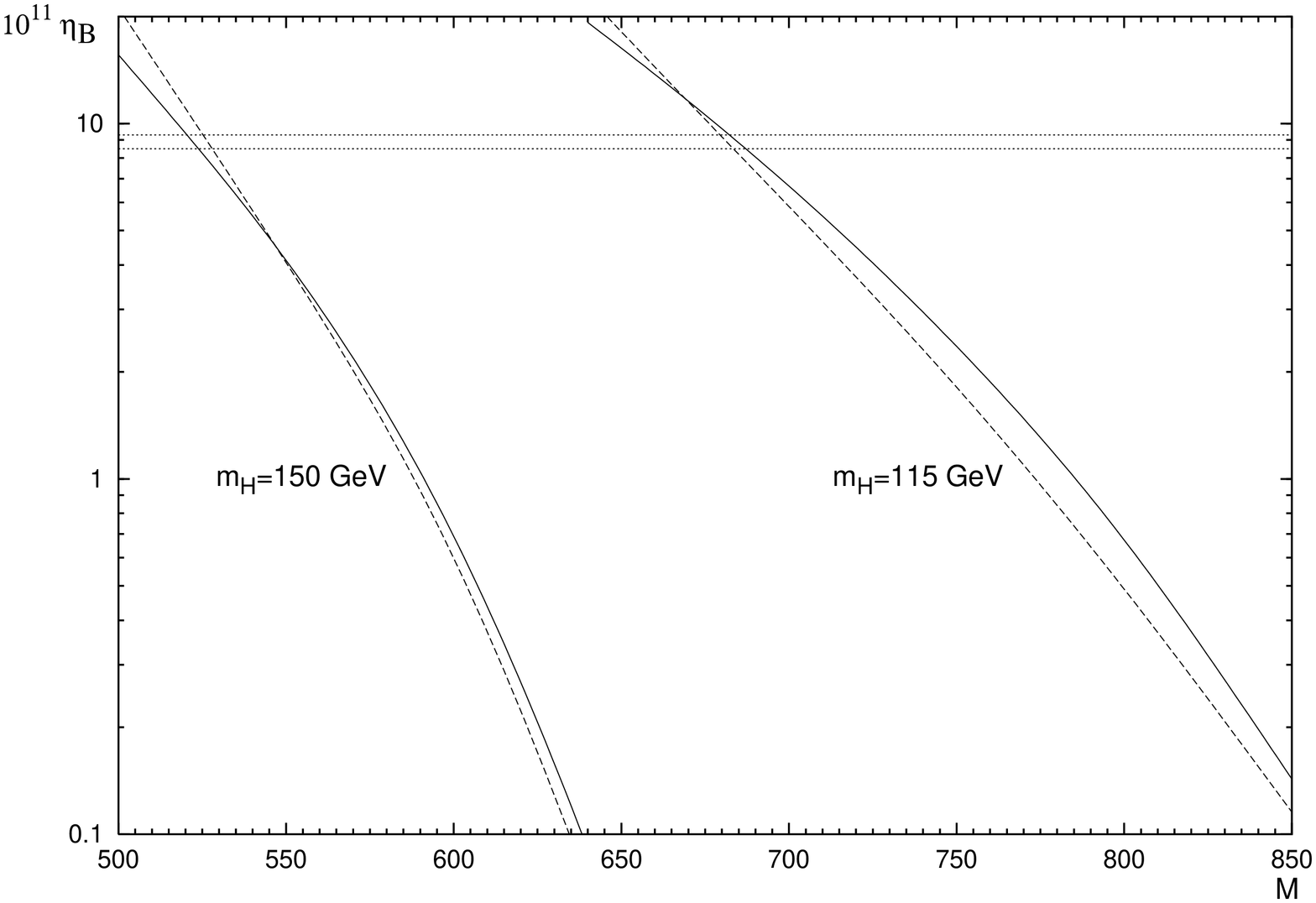,width=80mm,angle=270}
\caption{The baryon asymmetry in the SM with low cut-off for two
  different Higgs masses as a function
of $M$ (in units of GeV) for $v_{\rm w}=0.01$ (solid) and $v_{\rm w}=0.3$ (dashed). The horizontal lines indicate the errorband of the
measured value.
}
\label{model}
}
Let us finally relate the bubble wall parameters to our SM with low cut-off.
In the fig.~\ref{model} we present the baryon asymmetry in the model
for $m_H=115$ and 150 GeV as a function of the cut-off scale $M$. For every 
value of $M$ we compute the strength of the phase transition and the bubble
width. We consider one small $v_{\rm w}=0.01$ and one large wall velocity
$v_{\rm w}=0.3$. We take again a maximal CP violating phase 
$\sin\varphi_t=1$ and $x_t=1$. As expected the baryon asymmetry grows
rapidly as we lower $M$. At the very lowest values of $M$ the wall
thickness becomes of order $1/T$ (see fig.~\ref{LW}) so that our WKB approach
ceases to be reliable. Moreover, the bubble walls may become relativistic in this
regime, and diffusion of charges into the symmetric phase may no longer
be efficient. One can see that $\eta_B$ depends only mildly on the wall
velocity.

Nevertheless, independent of the Higgs mass we have chosen,
we can generate the observed baryon asymmetry (\ref{eta}) without
amplifying the CP violating dimension-6 operator (\ref{xij}). This requires
the phase transition to be sufficiently strong, i.e.~$\xi\gsim1.5$. At this
strength of the phase transition our computation of the baryon asymmetry
is still under control.
For smaller values of $\xi$ we have to take $x_t>1$. 
For instance, in the case of $m_H=115$ GeV and $\xi=1.1$ ($M=825$ GeV), we could
use $x_t \sim 40$ which is barely consistant with the bound from the neutron
EDM discussed in section 3.
Thus the SM with low cut-off can
account for the observed baryon asymmetry in a wide range of the model
parameters, without being in conflict with constraints from flavor and CP 
violation. 

At the end of this section let us briefly comment on the impact of
sources from the bottom quark and the tau lepton. In case of the bottom
the source term is heavily suppressed by $(m_b/m_t)^4\sim10^{-7}$.
The tau lepton is more relevant because of its larger diffusion constant
of about $380/T$ \cite{JPT1}. Still its contribution is about $10^5$ times
smaller than that of the top quark and can be safely neglected.

%
%
\section{Conclusions}
We have investigated the electroweak phase transition and baryogenesis
in the standard model augmented by a dimension-6 Higgs self interaction. 
Taking the suppression scale of this operator to be $M\lsim1 $ TeV, the EWPT
becomes first order, without introducing new degrees of freedom in the model.
In addition to the Higgs mass only the parameter $M$ enters the computation
of the phase transition. There is no relevant bound on the $\phi^6$ interaction.
However, dimension-6 operators involving for instance gauge fields, which are
also expected
to be present in a general effective field theory, have to be tuned at the
level of $10^{-2}$ in order not to spoil the electroweak fit \cite{GSW04}.

Requiring $M>500$ GeV the phase transition is strong enough to prevent baryon
number washout for Higgs masses up to 170 GeV. In our analysis we have used
the one-loop thermal potential. The phase transition is somewhat weaker 
than found in ref.~\cite{GSW04}, where only the thermal mass part of the
one-loop correction was taken into account. We have checked that the dimension-6
operator does not spoil the loop expansion of the effective potential.
We have computed the wall thickness which turns out to vary in a
wide range from about $2/T$ to $16/T$. As $M$ becomes smaller
the EWPT becomes stronger and the bubble wall thinner. For very small
cut-off scales the symmetric phase becomes metastable. 

A dimension-6 operator involving the Higgs field and the top quark
provides a new source of CP violation. It induces a complex phase
in the top quark mass which varies along the phase boundary.
We discussed that this operator is consistent with present bounds
on EDM's and flavor violation for $M\gsim 300$ GeV. 
However, it may leave a detectable signal in forthcoming experiments.  

As a result of the varying phase, top quarks and anti-top quarks experience
a different force as they pass through the bubble wall. We treat the
system in the WKB approximation, 
expanding in derivatives of the background Higgs field. 
Our approach is valid for a large fraction of the parameter space of the
model. It will break down for very small values of the cut-off scale $M$, where
the bubble width becomes of order the inverse temperature. The
CP violating force enters the transport equations which describe
the hot plasma. Carefully expanding in derivatives of the wall profile,
we find novel source terms which enhance the generated baryon
asymmetry. They are especially relevant for slow bubble walls
and dominate over the known source terms for large values
of the particle mass. Because of these properties they should play
a prominent role in the MSSM, where the top quark is
replaced by the charginos.  

In the model considered, the observed baryon asymmetry can be explained
for natural values of the parameters. The phase transition should be
somewhat stronger than required by the washout criterion. If the EWPT 
is not that strong, the coefficient of the CP violating dimension-6 operator 
has to be taken larger than one, which is compatible with experiments. It would
also be interesting to study the impact of other CP violating operators, such as
the one discussed in ref.~\cite{DHSS91}, which has been ignored in our study.
 
With a low cut-off the model is expected to lead to non-standard signals in
flavor physics, such as EDM's and flavor changing neutral currents, which
can be tested in future experiments.
The LHC will be able to directly test the physics at the cut-off scale.
If the general cut-off scale is in the multi-TeV range and the 
$\phi^6$ interaction is anomalously large, the model could still be
identified by its non-standard Higgs self couplings. However, the 
required precision will probably take a linear collider \cite{GSW04}.

In conclusion, the standard model with low cut-off provides the missing
ingredients for electroweak baryogenesis: a strong phase transition and
additional CP violation. Moreover, its simple structure makes it an ideal 
laboratory to refine the computation of the baryon asymmetry.

\section*{Acknowledgements}
We thank Dominik St\"ockinger, Steffen Weinstock and Oscar Vives
for helpful dicussions. The work of D.B., L.F., and M.S. was supported by
the DFG, grant FOR 339/2-1.


\end{document}